# Spontaneous and mass-conserved formation of continuous Si frameworks

**Authors:** K.Ogata[1,2], † D.-S.Ko[1], †* C. Jung[1], JH. Lee[1], SH. Sul[1], H.-G.Kim[1], JA. Seo[1], J.Jang[1], M. Koh[1], KH. Kim[1], J.H.Kim[1], I.S. Jung[1], M. S. Park[1], K.Takei[1], K.Ito[3], Y. Kubo[3], K. Uosaki[3], SG. Doo[1], S.Han[1], JK. Shin[1], and S.Jeon[1]*

†These authors equally contributed to the work
*Corresponding authors: seongho.jeon@samsung.com , ds02.ko@samsung.com

**Affiliations:**
1. Samsung Advanced Institute of Technology (SAIT), Samsung Electronics, Samsung-ro 130, Suwon, Gyeonggi-do, 16678, Korea
2. Samsung Research Institute of Japan (SRJ), Samsung Electronics, 2-1-11, Senba-nishi, Mino-shi, Osaka-fu, 562-0036, Japan
3. C4GR-GREEN, National Institute for Materials Science (NIMS), 1-1 Namiki, Tsukuba, Ibaraki, 305-0044, Japan

**Abstract**: (150 words)
Controlled formation of porous silicon has been of primary importance for numerous landmark applications such as light emitting sources, sensors, actuators, drug delivery systems, and energy storage applications. Frequently explored methods to form the structures have long relied on selective etching of silicon, which still stands as the most controllable and reliable methods to highlight essence of the applications. Here, we demonstrate an unprecedented approach to form silicon framework, which is spontaneously formed with atomistic arrangement of silicon without gravimetric loss via single electrochemical (de)alloying with lithium. Carefully controlling bare crystallinity of Si and composite/electrode designs, we reveal that the key prerequisite to forming the structure lies in using unique dealloying dynamics of crystalline–amorphous phase transformations at room temperature. Using the feature, we clearly highlight that commercially available nano-structured silicon particles can abruptly yet uniformly transform into continuous sub-2 nm spherical silicon frameworks with size-tunable pores.

Electrochemical dealloying of metallic compounds enables controlled formation of a variety of bicontinuous porous structures, (*1*) which is of great interest for numerous applications such as sensors, (*2*) catalysis, (*3*) actuation, (*4*) and drug delivery system (DDS). (*5*) For the system that lattice diffusion of the metal is slow (such as Au–Ag and Cu–Au), (*1*) the formation mechanism of the structure can be attributed to complicated interplay of the selective dissolution of the most electrochemical-active element and surface diffusion; known as percolation theory. (*6*) Analogous structures can be also achieved for the other alloy systems in which solid-state mass transport is significant and the dealloying reactions are governed by Nernst energetics (such as Li–M: M= Sn, Pb, Cd, Ge and Bi). (*7*) In such system, the percolation theory is not necessarily the prerequisite, but vacancy-mediated bulk diffusion can also be a driving-force to form the pores. The structure becomes bicontinuous structures, consisting of dendritic voids and ligaments, under mixed control of mass transport and charge transfer, (*1*) while Swiss-cheese-like voids may evolve under Kirkendall voiding.(*8*)

For silicon, the controlled formation of porous nano-structures (*9-11*) has been of paramount importance owing to various landmark applications represented by light emitting sources as well as the above-mentioned applications. (*11-13*) The fabrication methods, dating back in 1950's,(*9, 10*) have historically relied on selective etching processes, such as electrochemical anodization (*9, 10, 12, 14, 15*) and (metal-assisted) chemical stain etching. (*16-18*) Further, recent studies have shown a few different methods such as bottom-up synthesis, (*19*) $SiO_2$ reduction by Mg, (*20*) and dissolution of alkali metal from Li– (Na) –Si composites. (*21*) However, controlled formation of the porous structures using single alloying electrochemical dealloying has not been clearly reported, despite a few highlights on gradual coarsening/pulverization of Si or incidental cavity formation via electrochemical Li insertion/extraction in Si-based anodes for Li-ion batteries over cycling. (*22-25*) For the (de)alloying with lithium at room temperature, c-Si is gradually converted into amorphous-Li$_x$Si (a-Li$_x$Si). This is followed by abrupt phase transformation into c-Li$_{3.75(+\delta)}$Si at low voltages at < 70 mV, presence of which is significantly affected by crystallinity of bare Si, electrode design, and electrochemical conditions. (*24, 26*) On dealloying, c-Li$_{3.75(+\delta)}$Si hysterically decomposes into a-Li$_{\sim1.1}$Si at 430 mV under a quasi-two-phase reaction. (*27*) Notably, the phase transformation is typically recognized as a degradation cause in the anodes, mostly believed to simply pulverize, electrically isolate Si, and consequently degrade the anodes over cycling. Therefore, cycling conditions are often designed to avoid the formation. (*27*) In fact, attention has not been paid to quantitatively/qualitatively controlling formation of the phase transformation and correlating it with resulting microscopic Si morphologies.

Here, we present a new approach to controllably form nano-structured Si frameworks using single electrochemical lithium insertion/extraction at room temperature without gravimetric loss. Carefully controlling bare crystallinity of Si and composite/electrode designs, we reveal that the key prerequisite to forming the structure lies in using unique dealloying dynamics of crystalline–amorphous (c–a) lithium-silicide phase transformations. Using the feature, we clearly highlight that a variety of commercially available Si nanoparticles (SiNPs) can spontaneously and uniformly transform into continuous sub-2 nm spherical Si frameworks with size-tunable pores.

**Results**

*Material and electrochemical conditions*

Two different types of Si nano particles (SiNPs) with contrast crystallinity are used in this study; poly-crystalline (pc-) and single-crystalline (sc-)SiNPs. The different defect population in the particles allows us to control the formation of c-Li$_{3.75(+\delta)}$Si at the end of alloying (see Material and Methods and Table S1). pc-SiNPs (Fig. 1A) are more susceptible to the phase transformation compared to sc-SiNPs and consequently enables abrupt amorphous–crystalline (a–c) Li–Si phase transformation (a-Li$_{3.75}$Si → c-Li$_{3.75(+\delta)}$Si). Notably, we confirm that almost all the Si turns into c-Li$_{3.75(+\delta)}$Si even in the 1$^{st}$ cycle (Fig. 1A). Therefore, the presence/absence of c-Li$_{3.75(+\delta)}$Si at the end of alloying can be sharply controlled by the capacity-cutoff depth of discharge (DOD) near 100% (Fig. 1B–D). In contrast, sc-SiNP (Fig. 1E) has more difficulty in forming c-Li$_{3.75(+\delta)}$Si in the 1$^{st}$ cycle regardless of current rates (Fig. S1) probably due to less defective nucleation sites in the 1$^{st}$ cycle. In the following few cycles, c-Li$_{3.75(+\delta)}$Si gradually emerges and eventually becomes dominant at the end of the 4$^{th}$ alloying (Fig. 1F–H). Accordingly, the c-Li$_{3.75(+\delta)}$Si presence/absence can be controlled by the cycle number (≤ 4$^{th}$) in sc-SiNP. Thus, it is possible to clearly examine how presence/absence of c-Li$_{3.75(+\delta)}$Si, starting with various Si crystallinity, can affect the morphology of dealloyed Si. SiNPs are blended with multi-wall carbon nanotubes (MWCNT) via spray-drying processes and shaped into secondary particles (see Material and Methods). The electrodes are made of active materials, binder, and conductive agent (see Material and Methods). The electrodes assembled in coin half-cells (2032-type) are cycled under constant current constant voltage (CCCV) cycling on alloying and constant current (CC) during dealloying (see Material and Methods).

*Structural analysis of Si framework*

First, the morphology of fully dealloyed a-Si using pc-SiNP is observed by scanning transmission electron microscopy-high angle annular dark field (STEM-HAADF). The DOD is parameterized from 70 to 100 % for pc-SiNP in the 1$^{st}$ cycle. Note that an observation point in the anode is fixed at the center and surface (~5 μm) of the electrodes to ensure consistency of the observation. When the anode is cycled under DOD70–90% to undergo amorphous–amorphous (a–a) volume changes, the morphology of dealloyed a-Si is spherical bulk (Fig. 1I) as usually reported in previous studies. (*27*) In contrast, when the anode undergoes amorphous–crystalline (a–c) transformation under DOD100% on alloying and the subsequent c–a structural changes on dealloying, the morphology becomes continuous frameworks (Fig. 1J). The framework structure typically consists of sub-2 nm wires (standard deviation, σ, ~0.4 nm) and ~3–4 nm pores (σ~0.8 nm), sustaining the original spherical fringe. The formation of the framework is universal regardless of positions in the electrodes (Fig. S2 and Fig. S3). The accompanied elemental mapping images using electron energy loss spectroscopy (EELS) clearly show that the frameworks consist of Si (Fig. S4). The bulk stripes (~10 nm) seen in the framework structures are probably derived from amorphised twins and stacking faults, which have more difficulties in being fully alloyed due to higher energy cost to break Si–Si bonding. A small portion of the framework is confirmed under DOD90% cycling condition. This is probably due to the inhomogeneity of Li concentration during alloying, through which a portion of the Si near the separator side is converted into c-$Li_{3.75(+\delta)}Si$ and yields a portion of the frameworks. The signature of this can be seen at a portion of the 430 mV peak in the *dQ/dV* profiles for DOD90% (Fig. 1D, detail of the Li–Si processes is summarised in Material and Methods). For sc-SiNPs, the morphology of dealloyed amorphous Si in the 1$^{st}$ cycle is spherical bulk, which transforms into the frameworks after the 4$^{th}$ dealloying. The bulk–framework transformation corresponds to gradual c-Li$_{3.75(+\delta)}$Si dominance from the 1$^{st}$ to 4$^{th}$ cycle on alloying, and the signature can be seen in the gradual prominence of the 430 mV process instead of broad processes at 300 and 550 mV in dQ/dV profiles in Fig. 1G.

X-ray absorption fine structure (XAFS) analysis is conducted for the electrodes made of pc-SiNPs. The extended X-ray absorption fine structure (EXAFS) is Fourier transformed to interpret environments in Si–Si correlations, which is analogous to the radial distribution function (RDF). The Si–Si tetrahedral nearest bonding in dealloyed Si can be quantified by integrating and normalizing the 2 Å correlation peak. Interestingly, the index for the fully dealloyed Si linearly decreases from DOD50 to 90%, however, it abruptly drops between DOD90 and DOD100 % (Fig. S5). The sudden decrease of the index suggests that the four-coordinated Si is abruptly converted into reduced-coordinated ones. This is in line with the structural change from bulk to frameworks (Fig. 1I–L).

Three-dimensional (3D) structural analyses are conducted using electron tomography reconstruction (Fig. 2 and Movie S-1–4). The error bar for the tomography imaging tabulated from Crowther criteria (*28*) is smaller than 0.3 nm. The sliced images in the directions of the XY and XZ planes clearly highlight that the pores are persistently segmented by the continuously and uniformly spread 2-dimensional frameworks. This is in contrast to the bicontinuous structures in the precedents. (*1, 7*) The porosity of these structures, tabulated by numerically reconstructing the images (Fig. S6), is ~40–50 %. Essentially, the Coulombic efficiency (CE) upon the bulk–frameworks structural changes for pc and sc-SiNP is above ~92 % (1$^{st}$ cycle) and ~99 % (2$^{nd}$–4$^{th}$ cycle), respectively. Moreover, the frameworks formation is confirmed using different electrolytes (Fig. S7) which do not include fluorinated additives and/or fluorinated Li-salts. These results suggest that the origin of the framework is attributed to neither Si pulverization by mechanical stresses nor Si dissolution by potentially formed hydrofluoric acid (HF). It is to be noted that the formation of the framework is universal using 30–250 nm Si nano-particles with very contrasting crystallinity, generic electrode composition, and various electrolyte compositions.

*Identifying formation process of Si framework*

Li–Si alloys at different potentials are observed by *ex situ* transmission electron microscopies (Fig. 3A) in order to specify the process that forms the framework. Sample preparation considering the metastability of the Li-Si phases and the susceptibility of the alloys to electron-beam exposure is considered in Material and Methods. At DOD50 % on alloying (Fig. 3B), the structure consists of a-$Li_xSi$ phase with remaining bulky c-Si core, with the diffraction pattern solely showing the signature of residual diamond cubic structure of Si. At DOD100 % on alloying (Fig. 3C), the structure becomes lumpy and mossy bulk with clear diffraction pattern from c-$Li_{3.75(+\delta)}Si$. At depth of charge (DOC) of 50 %, i.e. nearly in the middle of the quasi-two-phase reactions (the frameworks and bulk are co-present), we focused on bulky alloy with the c-$Li_{3.75(+\delta)}Si$ diffraction pattern (Fig. 3D). Subsequently, the electron beam is projected to the structure from a time of $t$=0 to 15 min and the structural change is *operando* recorded; the continuous electron beam exposure enables induced self-dealloying (confirmed in Movie S-5 and Fig. S8). At $t$=0, Li–Si alloy with c-$Li_{3.75(+\delta)}Si$ signature in diffraction pattern is still bulky (Fig. 3E, left). At $t$=8 min, the diffraction spot pattern nearly disappears, which corresponds to the clear emergence of the framework (Fig. 3E, middle). At $t$=15 min, the diffraction spot pattern completely disappears (Fig. 3E, right) with presence of the framework. This demonstration indicates that the formation is induced by decomposition of c-Li$_{3.75(+\delta)}$Si not only under the electrochemical processes but also thermodynamic-like conditions by electron-beam exposure. *Operando* XRD analysis (Fig. S9) is also conducted, which certainly highlights that the c-Li$_{3.75}$Si decomposition proceeds under quasi-two-phase reaction at 430 mV (co-presence of c- Li$_{3.75}$Si and a-Li$_{\sim1.1}$Si), in which the XRD reflection from c-Li$_{3.75}$Si is continuously present on the plateau and gradually decreases its intensity. We also conducted $^7$Li NMR spectroscopy to probe local Si environments over the structural changes (Fig. S10). Key *et. al.* and Ogata *et. al.* (*24*) showed that when c-Si is fully converted into isolated Si (in form of c-Li$_{3.75}$Si) and overlithiated Si (c-Li$_{3.75+\delta}$Si) at the end of alloying under DOD100%, they hysterically reforms larger Si clusters (a-Li$_{\sim1.1}$Si) on the quasi-two-phase plateau at 430 mV, skipping formation of small Si clusters (a-Li$_{2.0-3.5}$Si). In contrast, for a–a structural change, Fig. S8 shows that the isolated Si (in form of a-Li$_{\sim3.5-3.75}$Si) and small Si clusters at DOD80% gradually build up the tetrahedral bonding from small to larger clusters upon two broad processes at 300 and 550 mV. Combined these results together, it is suggested that a driving force of the framework formation can be attributed to abrupt structural rearrangement of Si atoms on the hysteretic quasi-two-phase reactions. This is distinguished from mechanism in the frequently explored porous structures. (*1, 9, 10, 12*)

*Electrochemical parameter space to govern the morphologies*

Since the decomposition of c-Li$_{3.75(+\delta)}$Si controls the formation of the frameworks, we parameterized the alloying/dealloying current rates so as to investigate how the crystallinity of c-Li$_{3.75(+\delta)}$Si at the end of alloying and dealloying current rates in the following c–a decomposition alter morphology of the frameworks. Fig. 4A shows that the grain size of c-Li$_{3.75(+\delta)}$Si, tabulated from its XRD reflection profiles at 10 mV, increases from ~14 to ~30 nm when a alloying current rate decreases from 0.5 C to 0.005 C (see Materials and Methods). After forming different grain sizes of c-Li$_{3.75(+\delta)}$Si under different alloying rates, the crystal structure is fully dealloyed under different rates. Size of the frameworks does not significantly change regardless of the dealloying rates (Fig. 4B). In contrast, while size of the pores is constant under lower current rates, it prominently increases when dealloyed above 0.1 C (Fig. 4C) despite almost constant CE over different dealloying rates (Fig. 4D); the constant CE trend takes smaller values with decrease of the alloying rates probably due to a longer electrolyte exposure time to Si surface and consequent more irreversible Li consumption. c-Li$_{3.75}$Si crystal size (~14–30 nm) at the end of alloying does not significantly alter the framework/pore size profiles over the given dealloying rates. Interestingly, such trend is in contrast to the previously reported dealloyed systems. (*1, 7*) These results suggest that the formation is not primarily governed by percolation theory and surface smoothing. (*7*) These morphological evolutions on dealloying are schematically summarized in Fig. 4E, accompanied with the corresponding STEM-HAADF images.

## *Discussion*

In the previous sections, we showed that the framework formation emerges only through the unique dealloying dynamics that abruptly emerges upon the c–a phase transformation. Here, we preliminary discuss how the pores/frameworks are formed. One notable feature of c-Li$_{3.75}$Si is that Li diffusivity abruptly jumps up by two orders of magnitude upon the phase transformation (a-Li$_{3.75}$Si → c-Li$_{3.75}$Si) and the extraction of excess lithium (c-Li$_{3.75+\delta}$Si → c-

Li$_{3.75-\delta}$Si) (*29*); ~$10^{-12}$–$10^{-13}$ cm$^2$ s$^{-1}$ in a-Li$_x$Si (1.5<x<3.75) while ~$10^{-10}$–$10^{-11}$ cm$^2$ s$^{-1}$ in c-Li$_{3.75(\pm\delta)}$Si. Since Si is almost immobile with respect to Li in the alloy, their diffusivity difference abruptly increases near x=3.75 in Li$_x$Si. The sudden change may trigger Kirkendall voiding. (*8*) Another noteworthy feature of c-Li$_{3.75}$Si is that huge stress is generated at the quasi-two-phase interface (c-Li$_{3.75}$Si/a-Li$_{x\sim1.1}$Si) due to the significant volumetric difference. (*24*) It is known that the inward/outward propagation of such interface in such finite curvature structures yields complex time-dependent stress profiles in the sphere. (*30, 31*) This can be even more complex upon the c–a phase transformation since the reaction may occur at multiple sites in polycrystalline Li$_{3.75(+\delta)}$Si; tabulated grain size of c-Li$_{3.75(+\delta)}$Si is ~14–30 nm (Fig. 4A) in ~50–400 nm spheres. Further, ductility in the Li-rich Si alloys significantly increases (plastic deformation typically occurs at x > 2 in Li$_x$Si) (*32, 33*) and consequently more easily induces cavity formation under tensile. (*32, 33*) Therefore, alternative reasoning for the origin of the pore/framework formation is that the tensile that exceeds a critical value to nucleate a cavity may generate solely through the c–a dealloying process. Further, forming the highly-ordered pores can relax the interfacial stresses and reduce gross internal stress, which lowers the formation energy of the resultant a-Li$_{\sim1.1}$Si and consequently drive the pore formation. (*34*) One or a combination of these explanations may rationalize the origin of the framework formation; more detailed investigation will be given elsewhere. Another important note is that since the size of the pores is almost constant at lower dealloying rates, the pore growth by Ostwald ripening (*35*) can be excluded despite presence of pores in the ductile environments. Regarding the prominent increase of the pore size at higher alloying rates between 0.1 and 1C, increased internal stress at the two-phase interface with the increased dealloying rates may expand the nucleated cavities and/or force them to agglomerate. (*36*)

**Data availability**

The authors declare that all data supporting the findings of this study are available within this article, its Supplementary Information files, or are available from the corresponding author upon reasonable request.

**Methods**

*Baseline active materials*

Active materials in the form of secondary particles are synthesized by conventional spray-drying method (B290 Mini Spray-dryer, Buchi). These secondary particles consist of defective poly/single-crystalline (pc/sc-)Si nano-powder (pc-SiNP, Stream, ~120 nm and sc-SiNP, Cheorwon Plasma Research Institute, ~30–50 nm), multi-wall carbon nanotubes (MWCNT, 15 nm, CNT Co. Ltd.), and polyvinyl alcohol (PVA, Sigma Aldrich, MW ~ 50 k). First, the components are dispersed in DI water, followed by 2 h of ultrasonication. The mixed SiNP/MWCNT/PVA ratio for pc- and sc-Si-based secondary particle is 80/10/10 (wt %). The dispersed slurry is then spray-dried with a two-fluid-type nozzle at an inlet temperature of 220 °C in 60 standard litre per minute (slm) $N_2$ flow, with subsequent thermal treatment at 900 °C for 5 h in an $N_2$ atmosphere (100 °C/h ramping rate), followed by sieving (< 32 µm) to remove larger secondary particles. The secondary particles are designed to form a porous structure for better wettability and ensure the accessibility of Li ions to Si surface. The wt% of Si in pc- and sc-SiNP based secondary particles is quantified by inductively coupled plasma spectroscopy (ICP, Shimadzu quartz torch, Nebulizer-Meinhard-type glass) to be ~87 and ~73 wt%, respectively (Table S 1). These processes create well-defined physical parameters for the secondary particles. The average secondary particles have a diameter of ~10 µm with specific surface areas (SSAs) of 120 and 39.5 m$^2$/g for pc- and sc-SiNP based secondary particles, respectively. SiNPs and MWCNTs are well entangled in the secondary particles to secure good electrical connections and buffer space to accommodate the volume expansion of Si.

*Electrode and cycling conditions*

The electrodes are made of 79 wt% spray-dried secondary particles, 20 wt% polyacrylic acid (Li-PAA, Hwagyong Chemical) as a binder, and 1 wt% Kechen Black as a conductive additive. The components are mixed in a planetary mixer (Awatorirentaro, Thinky) for 15 min at 1000 rpm. The slurry is pasted onto a 10-µm-thick Cu foil, and the mass loading level (weighed by a Mettler Toledo XP26 instrument, ±1 µg accuracy) for pc- and sc-SiNP based electrodes are 1.3 and 1.8 mg cm$^{-2}$ (~3.0 and ~3.3 mAh cm$^{-2}$), respectively. 2032-Type coin cells (Hohsen Corp.) in a form of Li-metal countered half cells are used for cycling cells in all the experiments. The electrolyte is 1 M LiPF6 in a 25/5/70 (vol%) mixture of fluoroethylene carbonate (FEC)/dimethyl carbonate (EC)/dimethyl carbonate (DEC) (LP 30 Selectilyte, Merck). A 10-µm-thick separator (Asahi, Celguard, 1-µm-thick Al$_2$O$_3$ coated on both sides) is used. In this study, we define the electrode specific capacity (mAh g$^{-1}$) by dividing the total capacity (mAh) by the weight of spray-dried secondary particles on the electrode, i.e. 79wt% of the total mass loading, consequently being ~3342 and 2612 mAh g$^{-1}$ for pc- and sc-SiNP based electrodes, respectively (Table S 1). The capacity of the latter becomes smaller than the former since sc-SiNP wt% in the secondary particle gets smaller upon spray-drying due to difficulty of embedding smaller particles (~30–50 nm, in this case) into the secondary particles, usually being suck into vent. Here, the electrodes are in principle cycled at 1C under constant current constant voltage (CCCV) mode on alloying and constant current (CC) mode on dealloying. Capacity profiles of

DOD100% are defined such that the reference electrode reaches a current limit of 0.01 C in the constant voltage (CV) domain at 10 mV on alloying. A target DOD is controlled by capacity-cutoff referring to the capacity of DOD100%. For pc-SiNP electrodes, the presence of defects (stacking faults and twins) may offer more nucleation sites to form c-Li$_{3.75(+\delta)}$Si. Consequently, c-Li$_{3.75(+\delta)}$Si dominantly forms in the 1$^{st}$ cycle, which is clearly confirmed by the characteristic sharp peak in the dQ/dV profiles and capacity-voltage profiles at 430 mV (Fig. S 1C,D). This is in contrast to sc-SiNP based electrodes, in which the phase at the end of the 1$^{st}$ alloying is dominated by a-Li$_{\sim 3.5-3.75}$Si and gradually c-Li$_{3.75(+\delta)}$Si formation gets prominent in the end of the 4$^{th}$ alloying. A signature of this in half-cells is confirmed in the slope-plateau at 300 (a-Li$_{\sim 3.5-3.75}$Si → a-Li$_{\sim 2.0}$Si) and 550 mV (a-Li$_{\sim 2.0}$Si → a-Si) on dealloying. (26) The cells are cycled in a commercial closed-system cycler (TOYO system, TOSCAT-3100 series). The internal temperature of the cycler is maintained at ~23 °C (±1 °C) during the measurements. The internal system is set to use a current acquisition pitch of about ~1 s. Through calibrations, the instrument can measure the current with an accuracy of ±0.0167% (167 ppm) in a range of 2–10 mA, which consequently secures CE accuracy of as small as ±~0.07%.

*Lithium-silicon electrochemical processes*

When alloyed under depth of discharge (DOD)100%, the process in pc-SiNP during the 1$^{st}$ alloying is dominated by a sharp peak at 100 mV (c-Si → c-Li$_{3.75(+\delta)}$Si; gradual alloying of the c-Si lattice into a-Li$_x$Si, with further transformations into c-Li$_{3.75}$Si and c-Li$_{3.75+\delta}$Si). On dealloying, which is initiated by a rather flat process up to 300 mV (c-Li$_{3.75(+\delta)}$Si → c-Li$_{3.75(-\delta)}$Si), the characteristic plateau at ~430 mV is dominant (c-Li$_{3.75(-\delta)}$Si → a-Li$_{<\sim 1.1}$Si; a signature of conversion of c-Li$_{3.75(-\delta)}$Si into a Li-substituted amorphous phase), being the hysteretic

crystalline–amorphous phase transformation. When cycled below DOD90%, the dealloying process is dominated by broad processes at 300 and 550 mV, being amorphous–amorphous structural changes. For the processes in sc-SiNP under DOD100%, the ones in the 1$^{st}$ cycle undergoes the amorphous–amorphous structural changes regardless of (de)alloying current rates, which gradually changes into the crystalline–amorphous ones in the following few cycles with emergence of c-Li$_{3.75(+\delta)}$Si at the end of alloying.

*Ex situ TEM imaging and HAADF tomography*

The electrodes are characterized using SEM by slicing with a Ga focused ion beam (FIB) at 5 keV acceleration (Helios Nanolab 450F1, FEI). Then, *ex situ* TEM analyses are carried out using a double-Cs-corrected Titan Cubed microscope (FEI) at 300 kV with a Quantum 966 energy filter (Gatan Inc.) and a probe Cs-corrected Titan 80-200 microscope (FEI) at 200 kV. To avoid sample contamination and reaction upon air exposure, a vacuum transfer TEM holder (Model 648, Gatan, Inc.) and transfer vessel for FIB (hand-made) are used. All samples are moved from the FIB transfer vessel to the vacuum transfer TEM holder in Ar-purged glovebox. To fully delithiate the electrodes, the half-cell potential is held at 1.5 V for at least 24 h, until the current is less than 0.001 C. To avoid potential structural differences with respect to electrode positions, the analysed section is always chosen at the centre of the electrodes between 0 to 5 µm from the surface. To observe alloyed electrodes, the cell is cycled until potentials reach a target value and held there until the current decays to less than 0.001 C. The cell is disassembled in Ar-purged glovebox, and the electrode is promptly washed with dimethyl carbonate (DMC) for 5 min and then dried under vacuum for 30 min. Subsequently, the electrodes are scraped onto a lacey-carbon TEM grid (Ted Pella, Inc). For STEM-HAADF tomography imaging, the sample holder is tilted from -64° to 66° with 2° acquisition pitch, acquiring 66 slices for the image reconstruction. The sliced images are 3-dimensionally reconstructed and visualized using commercial software (Inspect3D v2.1 and Avizo v6.3, respectively: FEI). To reduce the noise in the raw image dataset, low pass filter was applied during the image reconstruction. For Movie S3 and S4, a single particle is extracted from Movie S1 and S2. This image processing necessitates intentional boundary cutting, by which surface of the particles artificially become smooth. The error bar for the tomography imaging based on Crowther criteria is expected to be <0.3 nm. (*35*)

*Ex situ XRD*

For *ex situ* XRD measurements, the coin half cells are cycled at 1 C under different DOD controls until the probing points. Subsequently, the electrodes are slowly alloyed at different alloying rates and the potential is maintained at 10 mV for at least 24 h until the current decays to less than 0.001 C to stabilize the metastable c-Li$_{3.75(+\delta)}$Si. Ogata *et. al* (*24*) showed that relaxation of the metastable phase becomes sluggish (at least 10 h) when cycled in this manner. After the cycling, the coin cells are immediately disassembled in Ar-filled glovebox, sealed with airtight Kapton tape, and immediately transferred to the XRD instrument (Bruker, D8 Advance). The measurements are performed using Cu Kα radiation (1.54 Å). Each spectrum is acquired in the range of 5–80° (2$\theta$) for ~50 min. The c-Li$_{3.75(+\delta)}$Si (332) reflection peak is fit by the Voigt function using a free software (Fytik) to determine its full width of half maximum (FWHM). The error in the FWHM is estimated to be 0.05°, considering the data acquisition pitch of the instrument.

*Ex situ $^7$Li solid-state NMR spectroscopy*

MAS $^7$*Li* solid-state (ss-)NMR experiments are conducted on Bruker Avance III consoles; $^1$H Larmor frequency of 600.13 MHz (14.1T). Commercial Bruker double-resonance 2.5-mm MAS probes that allow spinning frequencies up to 35 kHz are used for all experiments. $^7$Li MAS NMR spectra (233.2 MHz) are acquired *ex situ* at a spinning rate of 15 kHz with π/2- (one-pulse) measurements with a 2.0 s last-delay duration over 64 scans. After the coin half-cells reach the probing points at 1 C under different DOD controls, the cells are cycled at 0.02 C until reaching the target potential and held there for at least 24 h, until the current decays to less than 0.001 C. The cell is then immediately disassembled in an Ar-filled glovebox, dried for at least 30 min under vacuum, and packed in the rotor for the NMR measurements. All the $^7$Li ss-NMR chemical shifts are referenced to 1 M LiCl (sol.) at 0 ppm as an external reference. Based on previous studies by Ogata *et.al*. and Key *et.al*., (*24, 36*) the $^7$Li resonances are linked to Li–Si local environments as the followings: 10–0 ppm corresponds to larger Si clusters and extended Si networks; 25–10 ppm to small Si clusters; 6–0 ppm to isolated $Si^{4-}$ anions including c-$Li_{3.75}$Si; and 0– -10 ppm to overlithiated crystalline phase of c-$Li_{3.75+\delta}$Si.

*Operando XRD*

*Operando* synchrotron-based XRD experiments in a transmission mode were performed at beamline BL15XU (SPring-8, Japan) using an X-ray of energy 19 keV ($\lambda$= 0.653 Å). Diffraction patterns were collected at angle intervals of 0.01° using six one-dimensional detectors. Each XRD profile was collected in the 2$\theta$ region of 1.63°–74.37° every 5 min. The scan time was 22 s, which is much shorter than the operating duration of the test cell. In order to fully form c-Li$_{3.75(+\delta)}$Si, the cell analysed is pre-cycled in form of a 2032 coin cell to reach CV potential (10 mV) on alloying and held at least 12 h until current decays less than 0.005 C. This is followed by de-assembling of the cell and re-assembling in-house cell (the same size as a 2032 coin cell) for the *operando* measurement. The cell with φ2 mm pinholes at center was assembled in an Ar-filled glove box by sandwiching the fully lithiated anode (diameter: 12 mm), Teflon spacer, washer, curved washer, separator and SUS-coated Li foil to prevent any unintended exposure to the ambient environment during dealloying process (Fig. S 10). For window for X-ray transmission, the cell is internally sealed with Al-coated Kapton windows at top and bottom parts. X-ray is illuminated perpendicularly through the pinhole of the coin cell (transmission mode, Fig. S 10).

*Ex situ XAFS*

*Ex situ* XAFS at the Si K-edge is measured at the BL-10 of the Synchrotron Radiation (SR) Center at Ritsumeikan University. The photon beam energy delivered to the samples ranges from 1000 to 2500 eV with a resolution of 0.5 eV or less. 2032-Type coin half-cells are cycled at 0.1 C under designated cycling conditions. To fully delithiate the electrode, the half-cell potential is held at 1.5 V for at least 24 h until the current decays to 0.001 C. The cells are then disassembled in an Ar-filled glovebox. The electrode is rinsed with DMC for 5 min, set on carbon-taped sample holders, loaded into an airtight vessel, and then transferred to the BL-10 chamber without exposure to ambient air. The vessel is immediately evacuated, and the samples are loaded into the measurement chamber with a vacuum level of 5 × 10$^{-8}$ Pa. Partial fluorescence yield (PFY) mode is adopted to measure XAFS over the EXAFS range for the Si K-edge, which enables effective elimination of the P K-edge absorption signal by energy selected fluorescence detection with a Si drift detector (SDD). Small amounts of residual P on the surface of the Si anode could not be completely removed even after rinsing (typically < 5at %) yet does not invalidate the obtained profiles. The total electron yield (TEY) is also simultaneously measured, and P is detected in the EXAFS region of Si. Using open source analysis software (Athena), we extracted EXAFS data for dealloyed amorphous-Si (a-Si) at the probing points from XAFS and Fourier transformed, making them equivalent to radial distribution function (RDF) profiles. Tabulating the coordination number of a-Si involves a few uncertainties, such as statistical EXAFS fitting errors, sample preparation reproducibility, and assumptions made during data analysis for the physical structures surrounding the absorber. Hence, here we integrated the 2 Å Si–Si correlation peaks to index the Si–Si tetrahedral environments in the dealloyed a-Si. Oxidation of the anodes in the ambient environments after cycling is probably minimized, as the amplitude of the 1A coordination peaks (Si–O correlation) in the Fourier transformed profiles is lower than ~0.3. The error in $A_{(2Å\ Si–Si)}$, originating from sample reproducibility and handling issues, is not large enough to invalidate the overall trend observed.

**Acknowledgments**

We thank Toshiaki Ohta, Koji Nakanishi, and Toyonari Yaji for their insightful advices for XAFS analysis at Ritsumeikan SR center (BL-10). We also thank Chullho Song, Yoshio Katsuya, Masahiko Tanaka, and Osami Sakata for supporting our *operando* XRD experiments at SPring-8 (BL15XU).

**Author contributions**

K.O. and D.-S.K. schemed the concept and designed the experiments. K.O. and T.K. designed and synthesized the active materials. K.O. prepared all the electrodes and acquired all the electrochemical data. D.-S.K. and H.-G.K. acquired the TEM and STEM-HAADF tomography images. S.H.S processed and reconstructed tomography data. I.S.J. and J.H.K. acquired and processed the NMR data. K.I and Y.K acquired and processed XAFS data. J.-H.K. acquired *ex situ* XRD data. J.J., J.S. and M.K. synthesized electrolytes. K.I., K.H.K., and C.J. acquired and analysed *operando* XRD data. M.K., J.J., H.P., W.C, SG.D., K.U., S.H., J.K.S., and J.S. provided insights on the experiments. All authors wrote the article.

**Additional information**

**Competing financial interests**

The authors declare no competing financial interests.

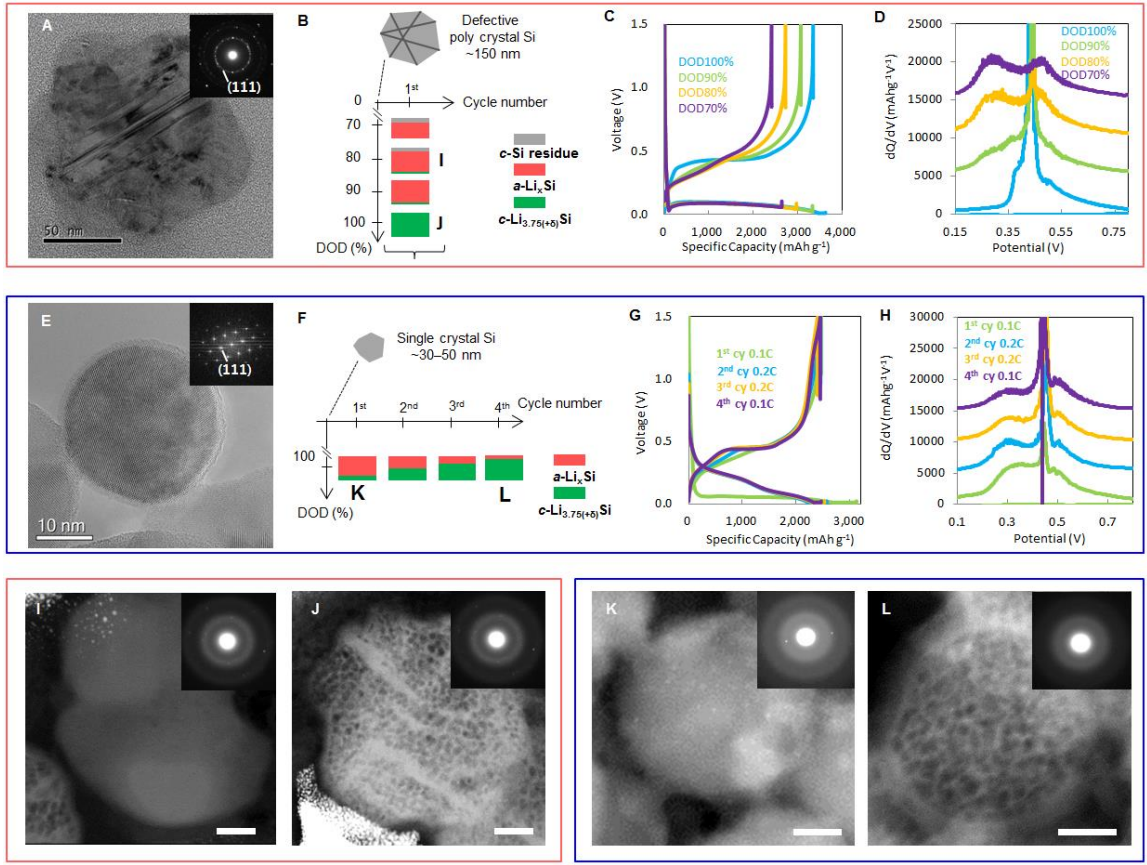

**Fig. 1 Transmission electron microscopic imaging for dealloyed Si cycled under different cycling protocols**

(A) TEM image of poly-crystalline Si with diffraction pattern at top-right inset. (B) Schematics that illustrate depth of discharge (DOD) control in the 1st cycle and corresponding lithium-silicide phase distribution at the end of each DOD. (C) Capacity-voltage profiles of poly-crystalline silicon for each DOD in the 1st cycle. (D) dQ/dV profiles for the corresponding capacity-voltage profiles in (C). (E) TEM image of single-crystalline Si with FFT image at top-right inset. (F) Schematics that illustrate cycle number control under DOD100% and corresponding lithium-silicide phase distribution at the end of alloying at each cycle. (G) Capacity-voltage profiles of single-crystalline silicon for each cycle under DOD100%. (H) dQ/dV profiles for the corresponding capacity-voltage profiles in (g). (I,J) Scanning transmission electron microscopy high angle angular dark field (STEM-HAADF) image for dealloyed poly-crystalline Si under DOD80% and 100%, respectively. (K,L) STEM-HAADF image for dealloyed single-crystalline Si at the 1st and 4th cycle, respectively. All the scale bars are 20 nm.

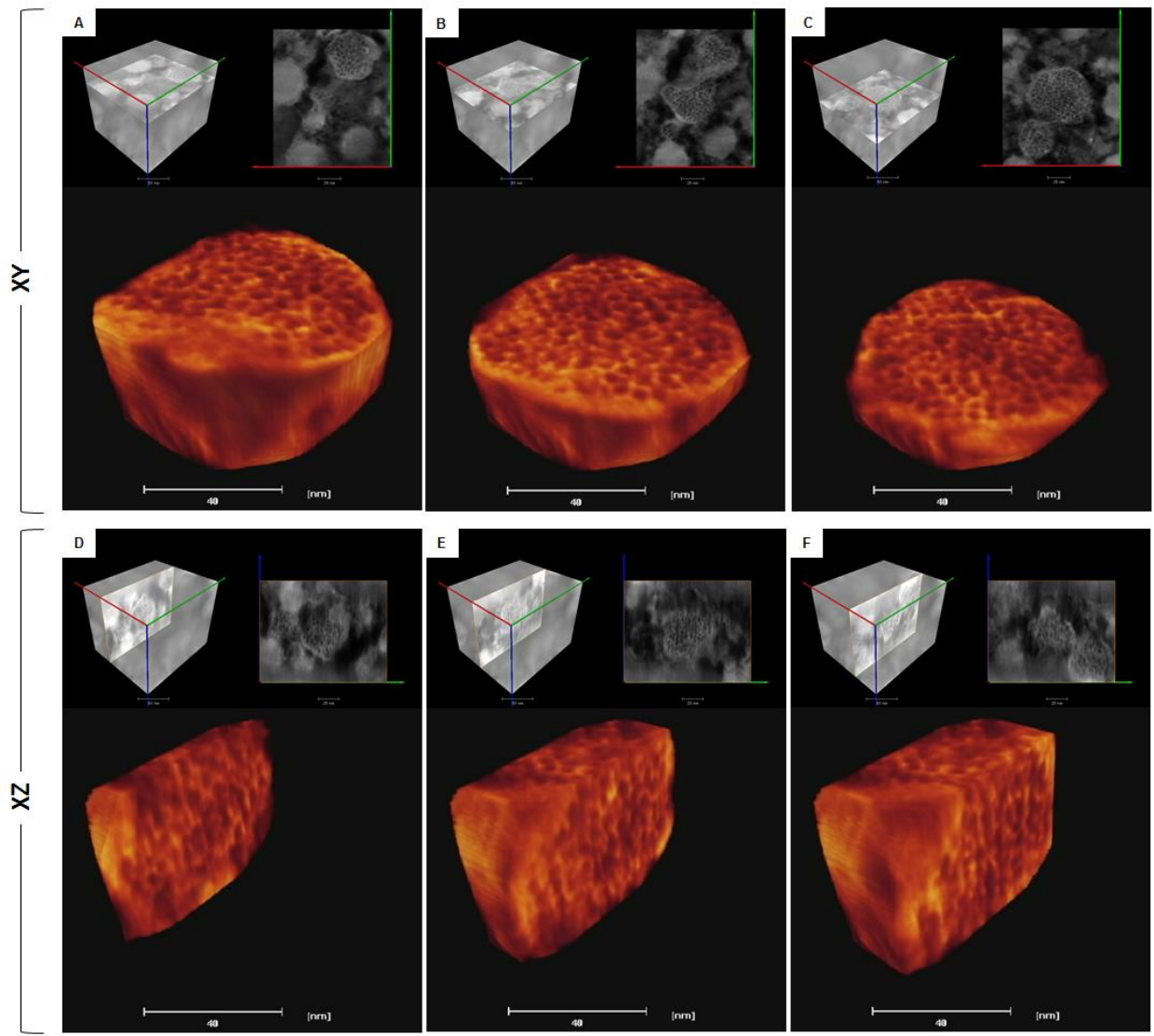

**Fig. 2 STEM-HAADF tomographic imaging of dealloyed single amorphous-Si particle**

(A–C) Sliced STEM-HAADF tomographic images by the XY plane for dealloyed single amorphous-Si particle fully through the crystalline–amorphous phase transformation. In the top part of each figure, X-Y-Z axes are indicated in 3-dimantionally reconstructed cubic-segment by green, red, and blue, respectively. In the lower part of each figure, topographically sliced single amorphous-Si particle after full dealloying is shown: brighter parts (orange) show Si, and the darker parts (black) highlight pores. (E–F) Identical particle sliced in the direction of the XZ plane. All the scale bars in the image show 40 nm.

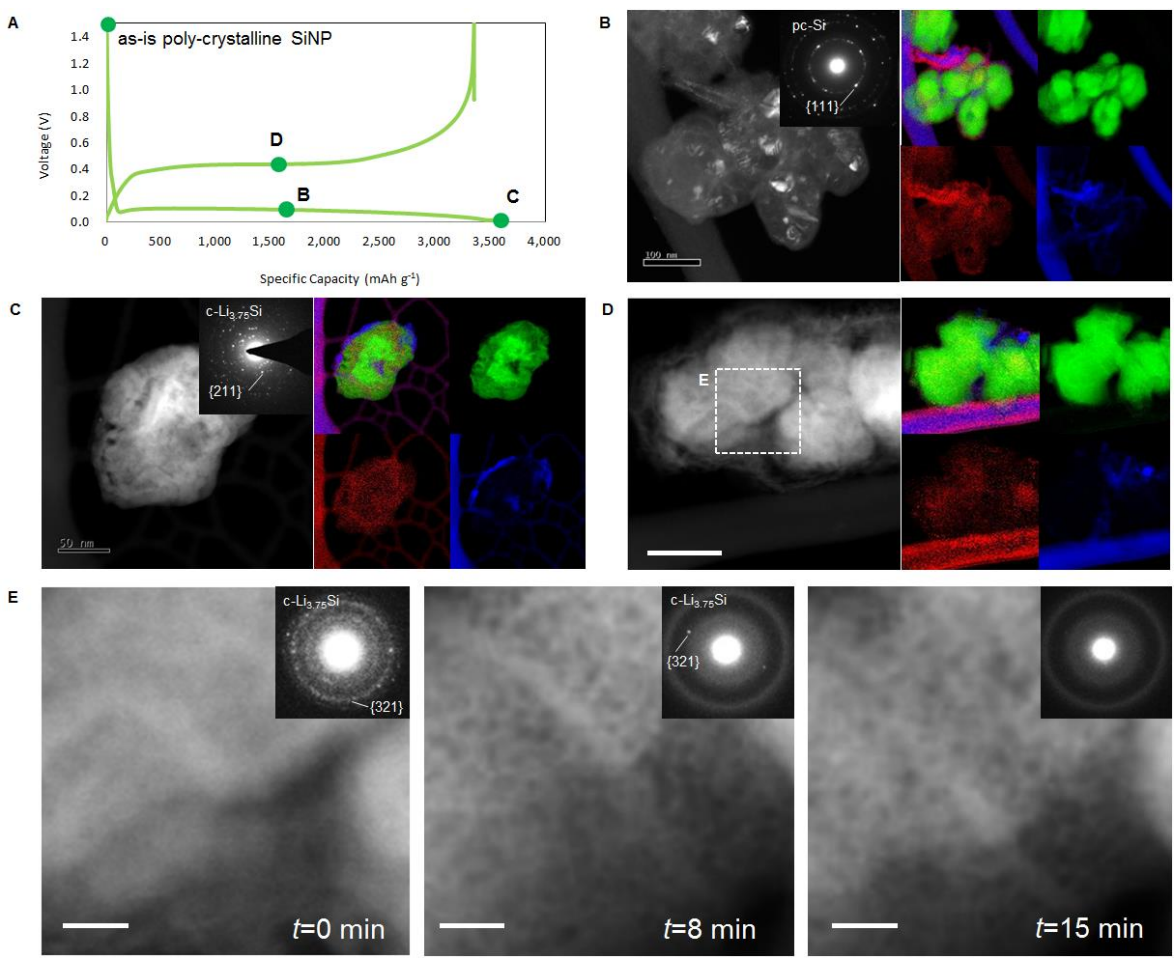

**Fig. 3 *Ex situ* transmission electron microscopy imaging for different (de)alloying potentials**

(A) Capacity-voltage profile for poly-crystalline silicon in the 1st cycle indicating ex situ transmission electron microscopic observation points on the profile. (B) STEM-HAADF image for depth of discharge (DOD)50% alloyed silicon particles with diffraction pattern at the top right and corresponding colored images show atomic EELS mapping of Si (green), Li (red), and C (blue). (C) STEM-HAADF image for fully alloyed (DOD100%) lithium-silicide with diffraction pattern of c-Li$_{3.75(+\delta)}$Si at the top right and corresponding colored images show atomic EELS mapping of Si (green), Li (red), and C (blue). (D) STEM-HAADF image for dealloyed silicon particle at state of charge (SOC)50% and corresponding colored images show atomic EELS mapping of Si (green), Li (red), and C (blue). (E) SOC50% lithium silicide is kept exposed to electron beam for 15 min (from left to right HAADF images in (E)). c-Li$_{3.75(+\delta)}$Si diffraction pattern at the top right in each image in (E) gradually disappears, which accompanies the emergence of Si frameworks. The scale bars in (B–E) are 100, 50, 50, 10, nm, respectively.

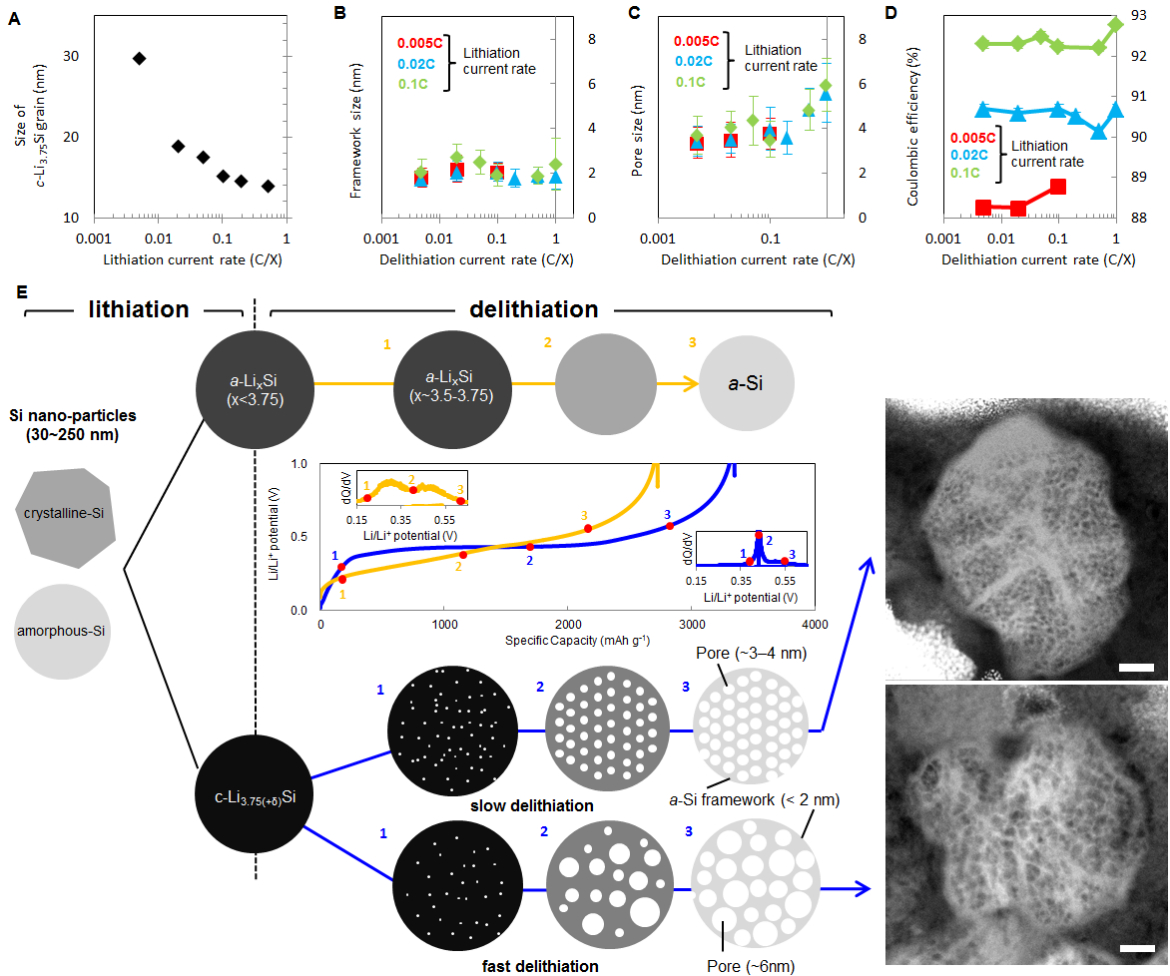

**Fig. 4 Morphology dependence of silicon frameworks on alloying/dealloying current rates.**

(A) Mean grain size of c-Li$_{3.75(+\delta)}$Si for different alloying rates in the first alloying at 10 mV (Li/Li$^+$). Size of (B) Si framework and (C) pores in dealloyed amorphous-Si after the 1$^{st}$ cycle for different alloying/dealloying rates. (D) Coulombic efficiency (CE) in the 1$^{st}$ cycle for different alloying/dealloying rates. (E) Schematics of morphology changes upon (de)alloying a single silicon nano-particle with/without presence of c-Li$_{3.75(+\delta)}$Si formation at the end of the 1$^{st}$ alloying. Representative capacity voltage profiles upon dealloying processes with absence/presence c-Li$_{3.75(+\delta)}$Si are shown in the middle of the schematics, which is further accompany corresponding dQ/dV profiles in top left and bottom right insets. The indicated numbers (1–3) in the profiles correspond to those on top left of morphologies on dealloying in the schematics. STEM-HAADF images for the corresponding schematics are also shown: all scale bars are 20 nm.

**Supplementary Information**

*Interpretation of the framework formation for battery applications*

Ogata *et. al.* showed that repeating c-Li$_{3.75(+\delta)}$Si formation/decomposition, which is often recognized as a degradation cause in Si-based anodes for Li-ion batteries, can non-linearly improve Coulombic efficiency (CE) and consequently minimize cumulative irreversible Li consumption in the anodes (available https://arxiv.org/abs/1706.00169v1). We attribute the origin of this to the spontaneous framework formation and to gradual interaction of the framework with the neighbouring ones over cycling. Further, we believe that this mass-conserved asymmetric volumetric change can lead to irreversible electrode thickness changes over cycling particularly when Si concentration in the anodes gets higher. Considering the high current density, e.g. >5–6 mAh/cm$^2$ in the state-of-the art Li-ion batteries, inhomogeneity of Li concentration perpendicular to the electrodes is inevitable upon realistic cycling condition. This leads to formation of c-Li$_{3.75(+\delta)}$Si in the electrodes near a separator side although the anode/cathode capacity loading ratio is typically designed to be <~1.05, i.e. maximum state of charge (SOC) in anode is less than 95%, being a-Li$_{<3.5}$Si. Therefore, it is to be noted that the insights in this work has significant correlation with pragmatic Si-rich Li-ion batteries.

*Optical properties of the frameworks*

We had attempts to detect photo luminescence (PL) from the Si frameworks, supposing that the 1.8 nm framework is not fully oxidized (SiO$_2$~9.3eV) and small enough to alter Si bandgap and potentially its structure. However, our preliminary data shows that PL from the structure was not prominent due to one of the following reasons. Firstly, we mix scratched powder of as-cycled electrode with epoxy and observed PL spectra. Hence there is a possibility that organic component in the electrode can disturb the PL. Secondly, the framework structure can be oxidized during preparation of the sample for PL analysis. We are further working on the way to extract potential presence of PL emitted from the structure.

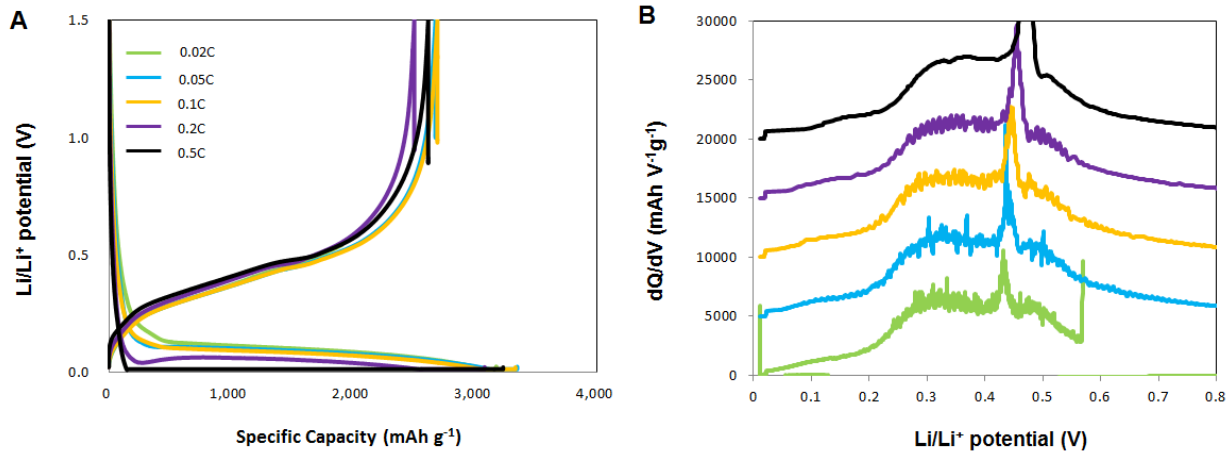

**Fig. S 1**

(A) Capacity-voltage curve of single-crystalline silicon in the 1st cycle for different (de)alloying current rates. (B) dQ/dV profiles on dealloying corresponding to the capacity-voltage profiles in (A).

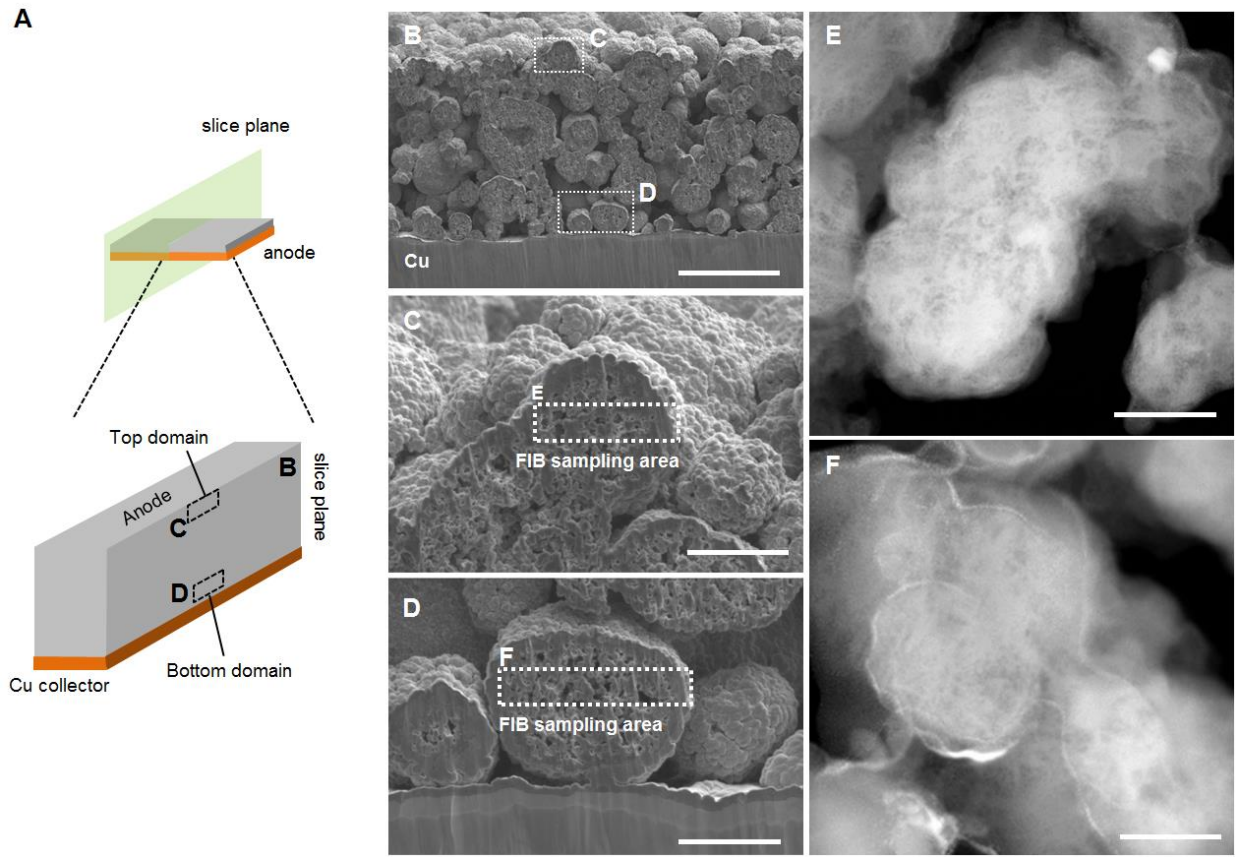

**Fig. S 2**

(A) Schematics of sliced electrode by Ar ion beam (IB-19520CCP Cross Section Polisher, JEOL, Ltd.). Different points of the electrode perpendicular to the electrode are indicated by dashed rectangles. (B) Cross-sectional SEM image of the sliced electrode. (C) Close-up SEM image of the electrode near surface. (D) Close-up SEM image of the electrode near Cu current collector. (E) STEM-HAADF image of dealloyed amorphous Si near surface of the electrode in (C). (F) STEM-HAADF image of dealloyed amorphous Si near Cu current collector in (D).

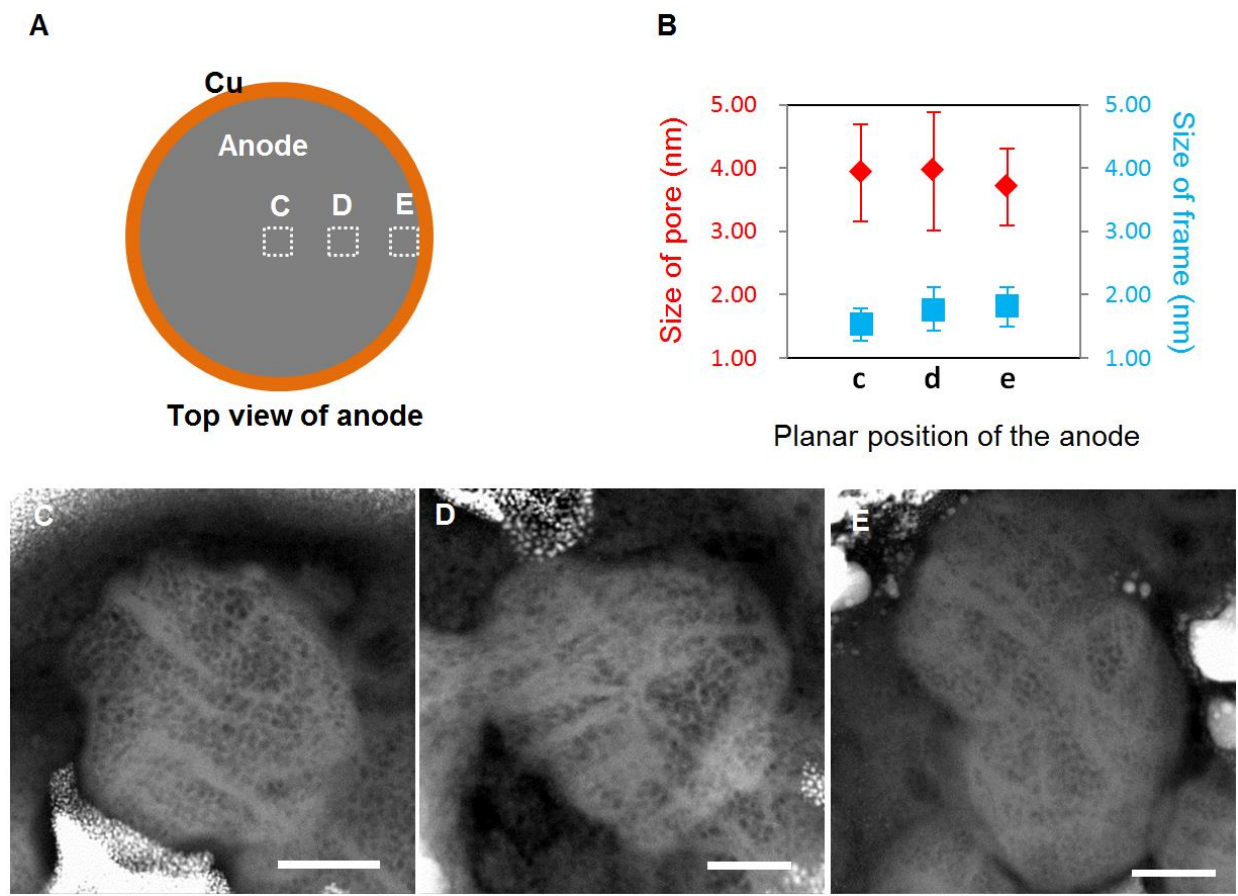

**Fig. S 3**

(A) Schematic of top-view of electrode, indicating 3 different planar positions on the surface of the electrode. (B) Size of Si frameworks and pores in dealloyed amorphous Si in the 3 different positions shown in (C-E). Cycling condition is CCCV on alloying and CC on dealloying at 0.1C. (C–E) STEM-HAADF images of dealloyed amorphous Si in the indicated positions in (A).

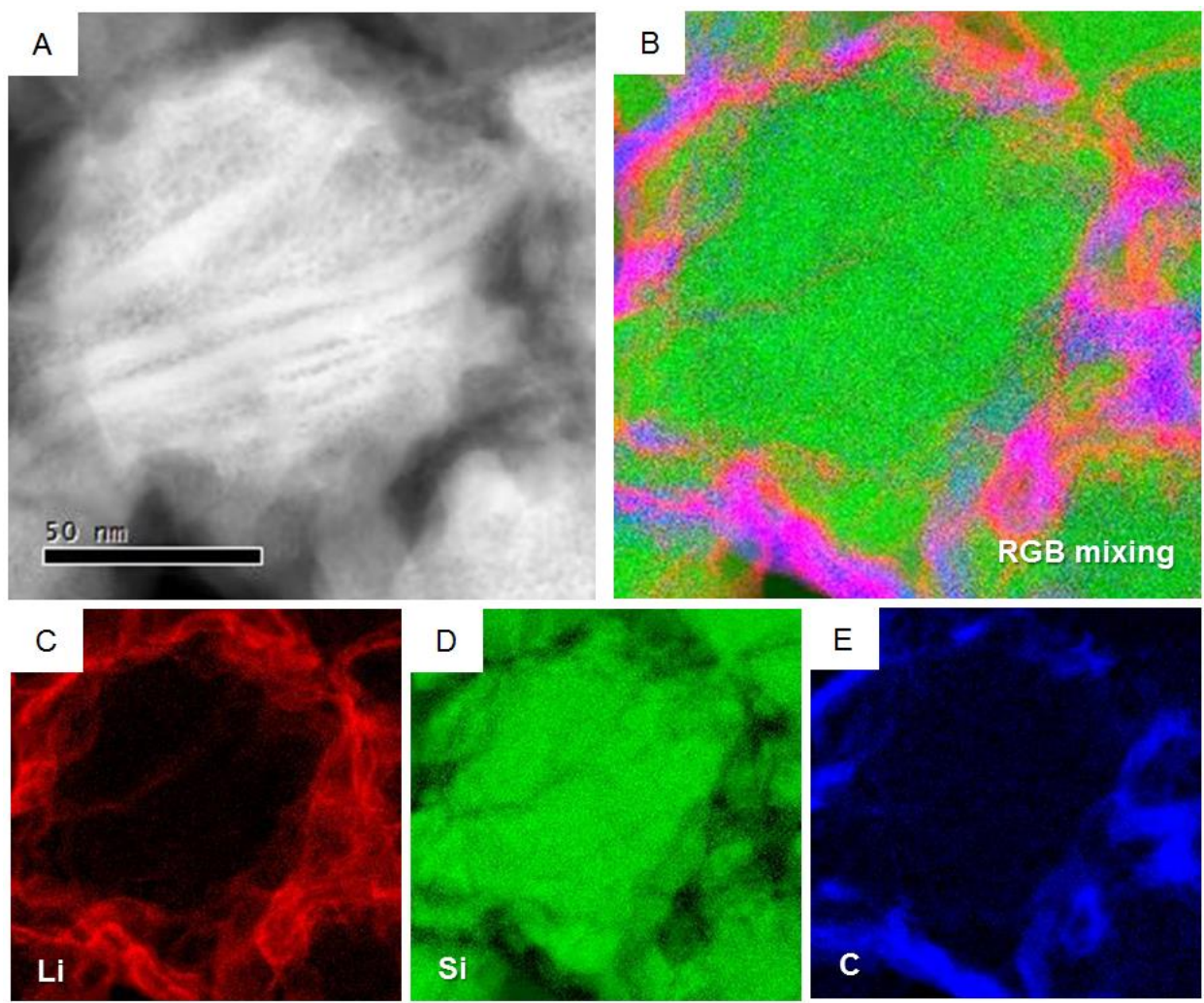

**Fig. S 4**

(A) STEM-HAADF image of dealloyed amorphous Si cycled under 0.1C CCCV on alloying and CC on dealloying. (B–E) Corresponding EELS atomic mappings of dealloyed amorphous-Si in the 1st cycle cycled under DOD100% at 0.1C. The mapping with red, green, and blue corresponds to Li, Si, and C, respectively.

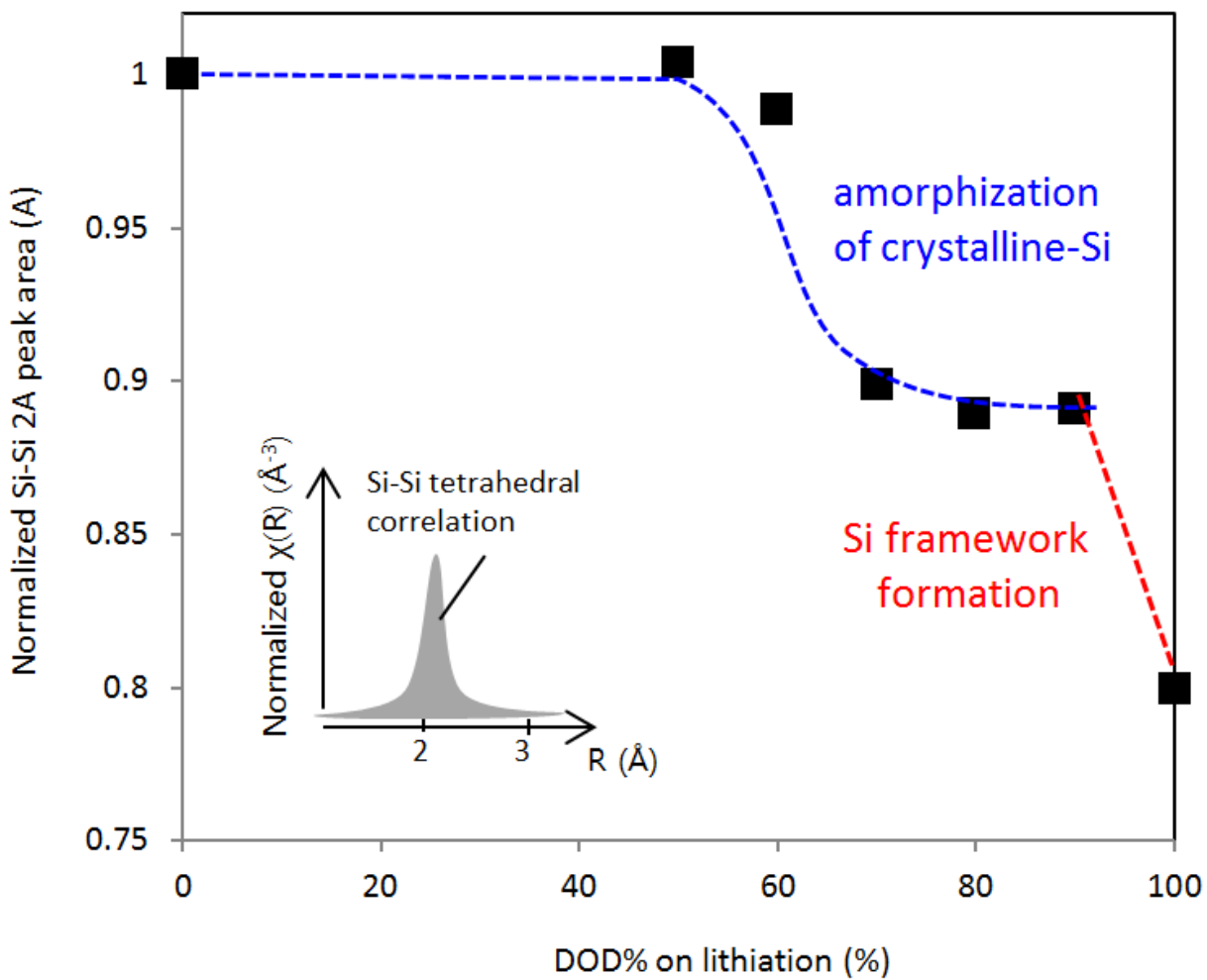

**Fig. S 5**

Plots of normalized integration of 2 Å Si–Si correlation peak in radial-distribution-function-equivalent profiles, acquired by Fourier-transforming Si EXAFS profiles in dealloyed amorphous-Si after the 1$^{st}$ cycle (indicated by schematics in the plots). The integrated area is named $A_{(2Å\ Si–Si)}$ and plotted for different DOD controls in the 1$^{st}$ cycle.

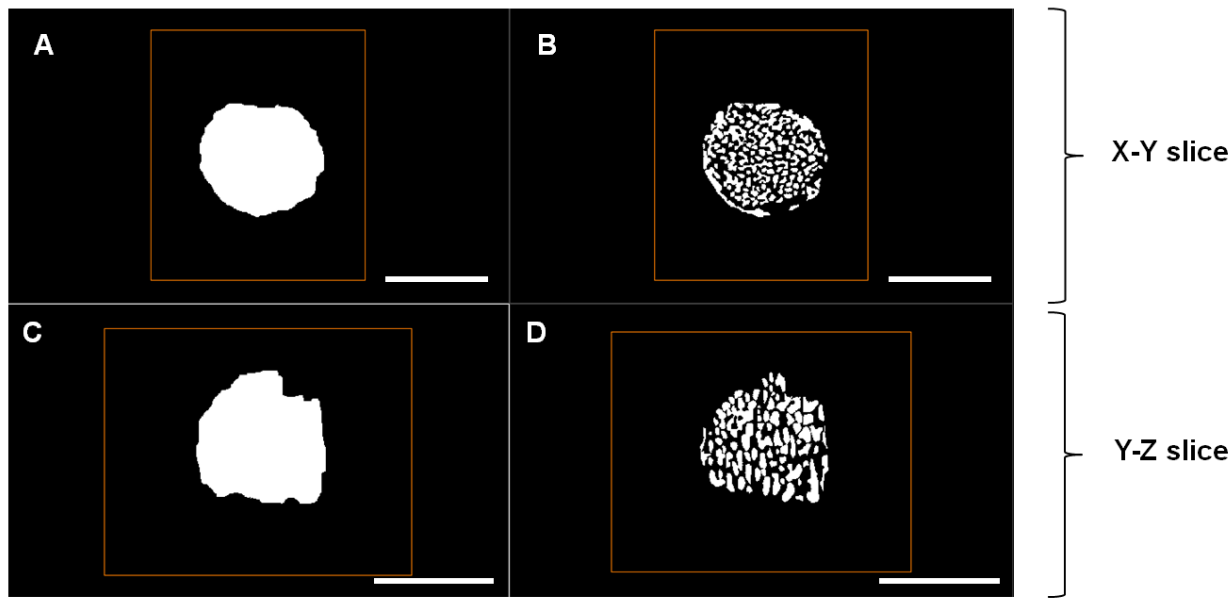

| | Image | Area ( pixel² ) | Porosity |
|---|---|---|---|
| A | xy particle | 33026 | 100 |
| B | xy void | 15969 | 0.483528 |
| C | yz particle | 26756 | 100 |
| D | yz void | 12985 | 0.485312 |

**Fig. S 6**

Planar projection of sliced dealloyed single amorphous-Si particle in different slicing planes for porosity calculations. XY plane projection of (A) whole area and (B) pores. YZ plane projection of (C) whole area and (D) pores. The table at the bottom shows area of the whole projection and pores in square of pixel with corresponding porosity. All the scale bars are 80 nm.

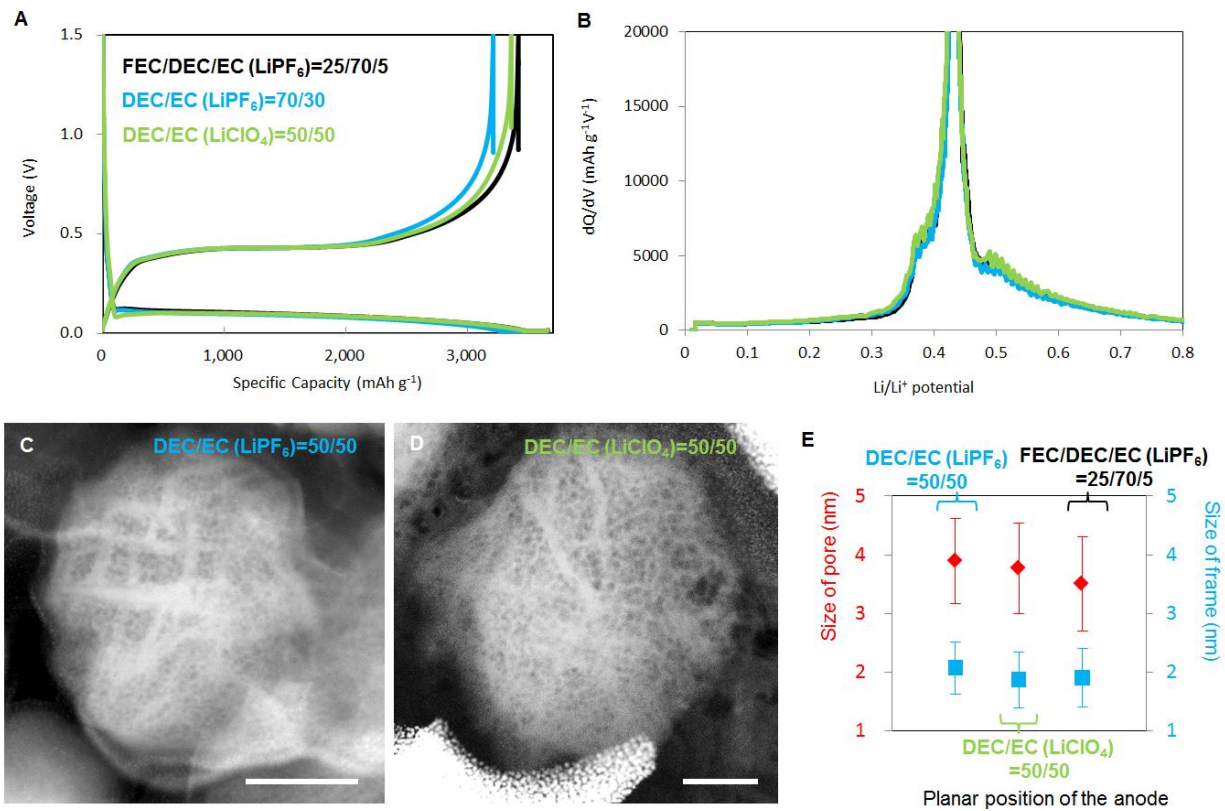

**Fig. S 7**

(A) Capacity-voltage curve of poly-crystalline silicon in the 1$^{st}$ cycle cycled at 0.1C for different electrolytes. (B) dQ/dV profiles on dealloying corresponding to the capacity-voltage profiles in (A). (C,D) STEM-HAADF images of dealloyed amorphous-Si for different electrolytes, composition of which is decribed at the top right part of each image. (E) Size of Si framework and pores in dealloyed amorphous Si in 3 different electrolytes. All the scale bars are 50 nm.

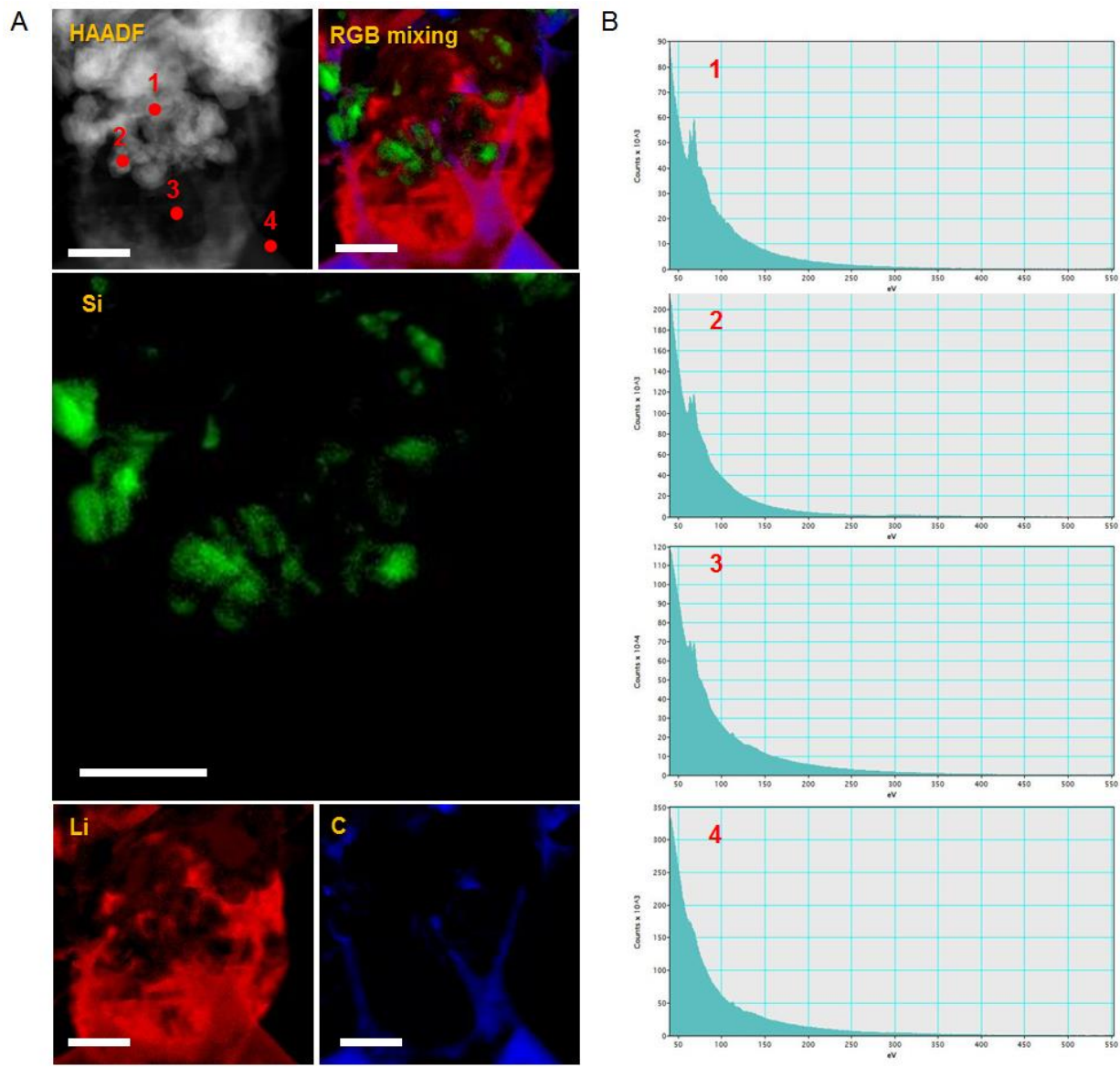

**Fig. S 8**

(A) STEM-HAADF images and corresponding EELS atomic mappings of self-discharged Li from c-$Li_{3.75}Si$ by consistent electron beam exposure. The mapping color with green, red, and blue corresponds to Si, Li, and C, respectively. (B) EELS spectra for different spots in (A). All the scale bars are 200 nm.

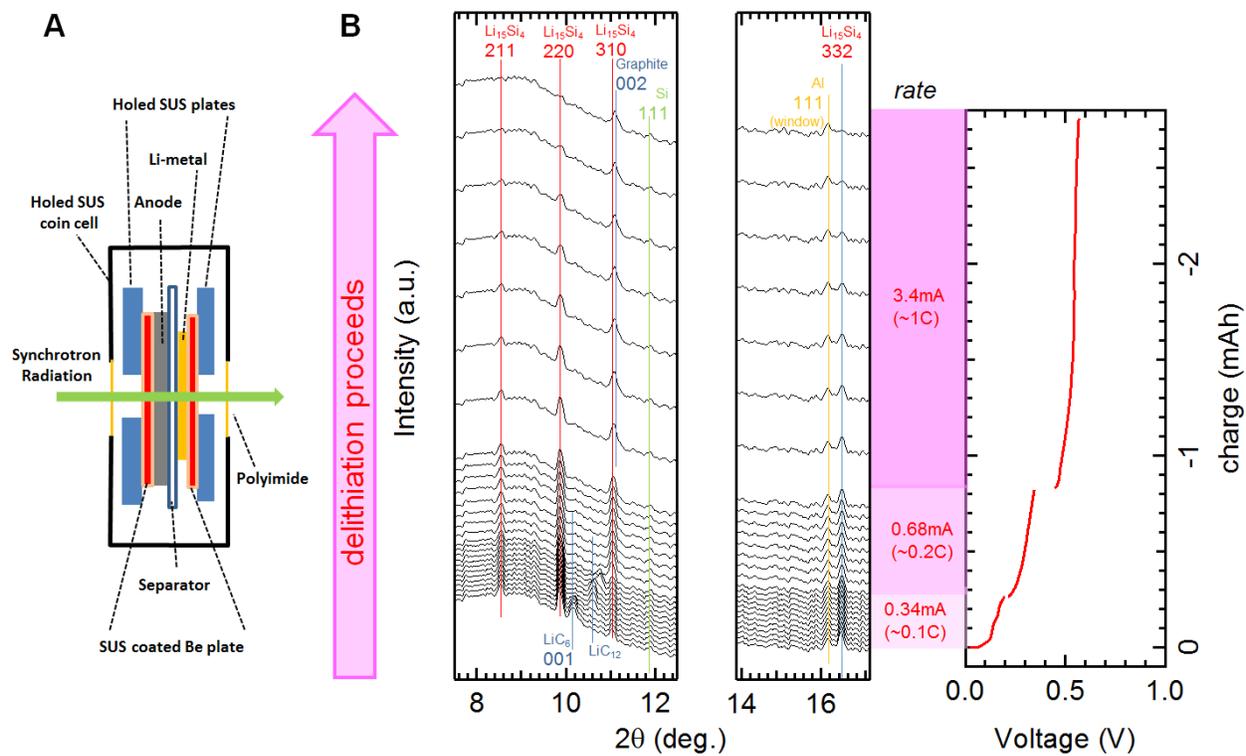

**Fig. S 9**

(A) Schematics of cell configuration for *operando* XRD measurements. (B) *Operando* XRD reflection profiles for the 1st dealloying process, which is linked to corresponding capacity-voltage profile.

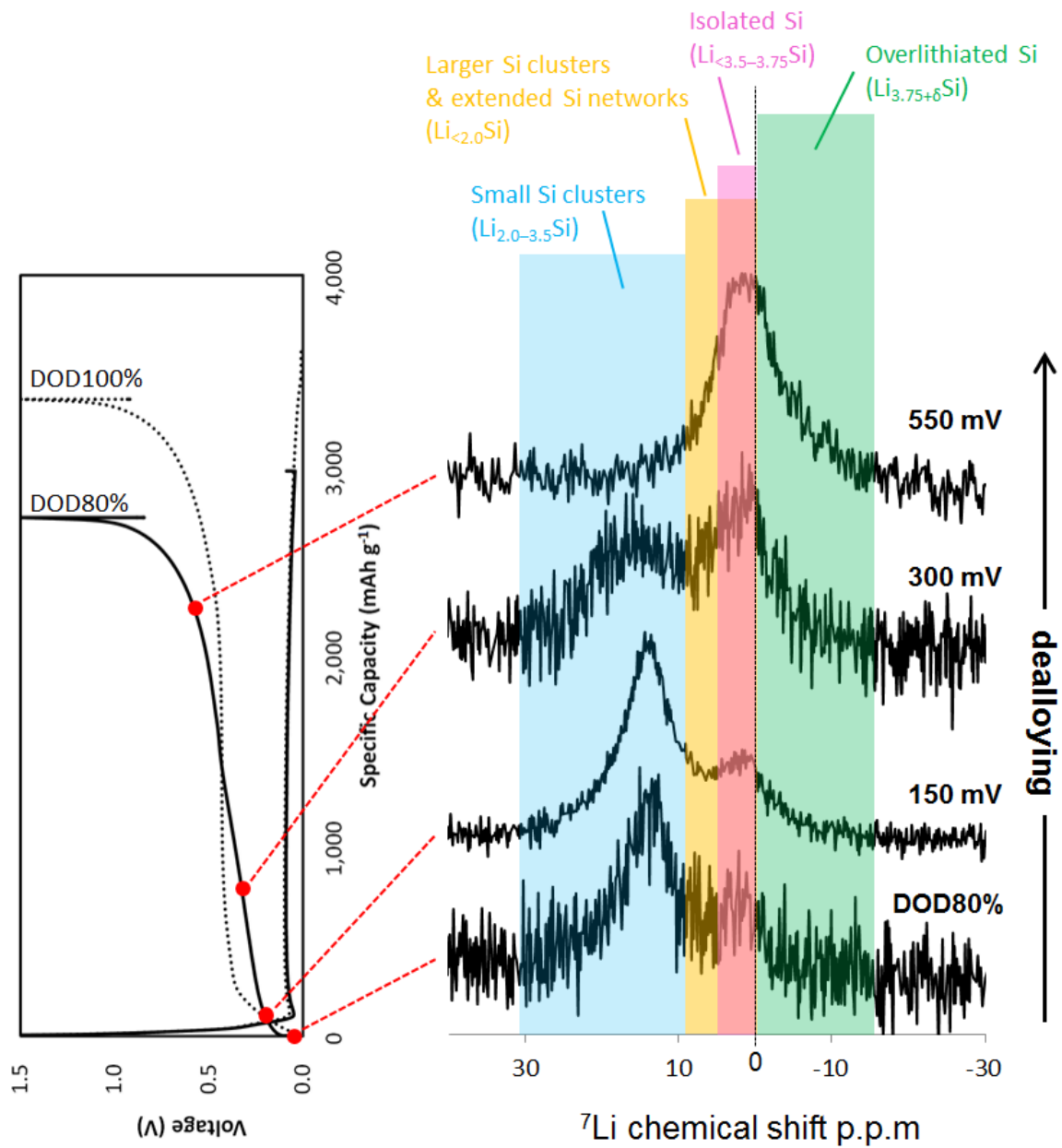

**Fig. S 10**

Spectra of *ex situ* $^7$Li NMR spectroscopy at different potentials on dealloying under DOD80% condition in the 1$^{st}$ cycle. The spectra are linked with the corresponding points on a capacity-voltage profile. The $^7$Li resonances are highlighted by yellow (10–0 ppm, larger Si clusters and extended Si networks), blue (25–10 ppm, small Si clusters), pink (6–0 ppm, isolated Si$^{4-}$ anions including c-Li$_{3.75}$Si), and green (0– -10 ppm, overlithiated crystalline phase, c-Li$_{3.75+\delta}$Si).

**Table S 1**

Physical parameters for secondary Si-C composites in a form of secondary particles consisting of single and poly crystalline Si.

| | | Single crystalline Si | Poly crystalline Si |
|---|---|---|---|
| **Si sources** | Mean size of Si sources (nm) | 30-50 | 150 |
| | Mean grain size of Si sources (nm) | 45 | 13 |
| **Secondary particle compositions** | SiNP (wt%) (via ICP) | 73 | 87 |
| | CNT (wt%) | 25 | 11 |
| | PVA (wt%) | 2 | 2 |
| **Secondary particle parameters** | Surface area of secondary particles (m$^2$/g) | 120 | 39.5 |
| | O$_2$ content of secondary particle (wt%) | 11 | 2~3 |
| **Electrochemical data** | Initial Coulombic efficiency (%) | 81 | 92.4 |
| | Theoretical capacity (mAh/g) (tabulated from Si wt% by ICP) | 2612 | 3342 |
| | Experimental capacity (mAh/g) | 2590 | 3350 |

**Movies S1**

Tomography 3D rendering movie: dealloyed amorphous-Si particles (originated from poly-crystalline Si particles) after the 1$^{st}$ Li insertion/extraction.

**Movies S2**

Tomography 3D rendering movie: dealloyed amorphous-Si particles (originated from single-crystalline Si particles) after the 1$^{st}$ Li insertion/extraction.

**Movies S3**

Single particle extracted tomography 3D rendering movie: dealloyed single amorphous-Si particle (originated from poly-crystalline Si particles) after the 1$^{st}$ Li insertion/extraction.

**Movies S4**

Single particle extracted tomography 3D rendering movie: dealloyed single amorphous-Si particle (originated from single-crystalline Si particles) after the 1$^{st}$ Li insertion/extraction.

**Movies S5**

*Operando* TEM movie of bulk Li metal formation through self-dealloying of Li$_x$Si composite induced by continuous electron beam exposure.